\documentclass[fleqn,twoside]{article}
\usepackage{espcrc2}

\usepackage{graphicx}
\usepackage[figuresright]{rotating}

\newcommand{\AmS}{{\protect\the\textfont2
  A\kern-.1667em\lower.5ex\hbox{M}\kern-.125emS}}
\newcommand{\be}{\begin{equation}}
\newcommand{\ee}{\end{equation}}
\newcommand\as{\alpha_s}

\newcommand\beqy{\begin{eqnarray}}
\newcommand\eeqy{\end{eqnarray}}
\newcommand\jpsi{J/\psi}
\newcommand\etac{\eta_c}
\newcommand\etab{\eta_b}
\newcommand{\aqed}{\alpha}

\title{Theoretical determination of $\etab $'s electromagnetic decay width}
\author{Nicola Fabiano\address[PG]{Perugia University and INFN, via 
A.~Pascoli I--06100 Perugia, Italy}}
\begin{document}

\begin{abstract}
We discuss the theoretical predictions for the two photon decay width of the 
pseudoscalar etab meson. Predictions from potential models are examined. It is 
found that various models are in good agreement with each other. Results for 
etab are also compared with those from Upsilon data through the 
NRQCD procedure.
\end{abstract}
\maketitle
\section{Introduction}
First evidence of $\etab$ was given in 2001 by ALEPH~\cite{ALEPH}. 
The search for $\etab$ is done in the decay channel into two photons. 
One candidate event is found in the 
six--charged particle final state and none in the four--charged particle 
final state, giving the upper limits:
$ \Gamma(\etab \to \gamma\gamma)\times BR(\etab \to 4 \textrm{ charged
particles}) < 48 \textrm{ eV , } $
$ \Gamma(\etab \to \gamma\gamma)\times BR(\etab \to 6 \textrm{ charged
particles}) < 132 \textrm{ eV .} $
This observation was the main motivation for a theoretical estimate of 
electromagnetic  decay width for $\etab$~\cite{JA}.
More recently CDF has done a search for $\etab$ in the process~\cite{CDF2K2}
$ \etab \to \jpsi \jpsi $
with 7 events where 1.8 events are expected from background.

This note is organised as follows: in Sect.~2 we shall compare the 
two photon decay width with   the leptonic width of the $\Upsilon$. Sect.~3
is devoted to the potential model predictions for $\etab \to \gamma \gamma$, 
with potential given by~\cite{ROSNER,IGI,NOI1}.
In Sect. 4 we show the predictions for $\etab$ decay widths, using
the  procedure introduced in~\cite{BBL} for the description of 
mesons made out of two non relativistic heavy quarks, by means of the Non
Relativistic Quantum Chromodynamics--NRQCD. 
In Sect. 5 we  compare these different  determinations 
of the  $\etab \to \gamma \gamma$ decay width 
together with a result based on a recent two-loop theoretical analysis of the 
charmonium decay~\cite{CZARNECKI}.

\section{Relation to $\Upsilon$ electromagnetic decay width}

\begin{figure}[h]
\begin{center}
\includegraphics*[angle=0,width=0.4\textwidth]{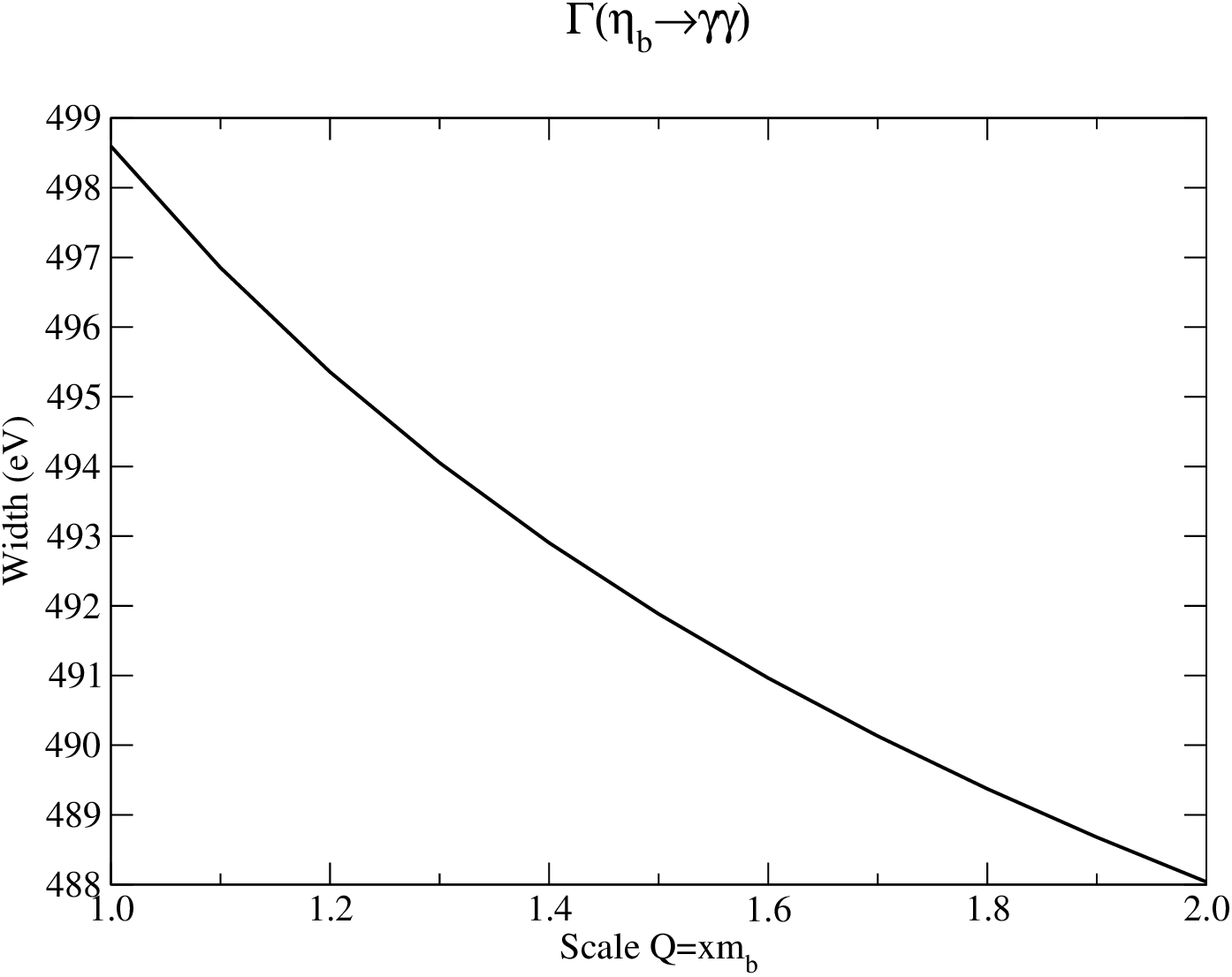}  
\caption{ The dependence  of the  $\etab$ decay width to $\gamma\gamma$ 
(in eV) is shown with respect to  the scale chosen for $\as$~in the 
radiative corrections. The fluctuation is of order 2\%\label{fig:oneloop}}
\end{center}
\end{figure}

We start with the two photon decay width of a pseudoscalar
quark-antiquark bound state in the singlet picture, generically written as 
\be \Gamma = NP \times P \ee
$NP$ being the nonperturbative part and $P$ the perturbative correction. The
nonperturbative part for a state with a given $\ell$
takes the form $NP \sim | d^{\ell} \psi(0)/dr^{\ell} |^2$. 
The wavefunction is obtained by a solution
of a Schr\"odinger equation with a suitable potential $V(r)$.
The perturbative expression is given by $P = F(\as(m_q))$ and comes from
the on-shell matrix element.
The two photon decay width of a pseudoscalar quark-antiquark bound 
state~\cite{VANROYEN} is given by

\be
\Gamma_B (\eta_b\rightarrow \gamma\gamma)=12
Q^4\alpha^2 4\pi \frac{|\psi(0)|^2}{M^2}\equiv \Gamma_B^P 
\ee
$M=2m_b+E_b$ being the mass of the state, $E_b$ the (negative) binding energy;
with first order QCD corrections \cite{BARBIERI}, which can be written 
as
\be
\Gamma(\etab\rightarrow \gamma\gamma)= \Gamma_B^P\left 
[ 1 + 
\frac{\alpha_s}{\pi} 
\left(\frac{\pi^2-20}{3} \right ) \right ]
 \label{eq:widsc1l} .
\ee
A first theoretical estimate for this decay width can be obtained by 
comparing eq.~(\ref{eq:widsc1l})  with the  expression for 
the electromagnetic decay for the vector state $\Upsilon$~\cite{MACKENZIE}, i.e.
\be
\Gamma_B (\Upsilon\rightarrow  e^+e^-)=4
Q^2\alpha^2 4\pi \frac{|\psi(0)|^2}{M^2}\equiv \Gamma_B^V 
\ee
and the one-loop complete formula
\be
\Gamma (\Upsilon\rightarrow e^+e^-)=\Gamma_B^V \left ( 1-
\frac{16}{3}\frac{\alpha_s}{\pi} \right )  .\label{eq:widvec1l}
\ee
Using the  expressions in eqs.~(\ref{eq:widsc1l}) and~(\ref{eq:widvec1l}) we 
can estimate the $\etab$ decay width from the the measured values of the
leptonic decay width of $\Upsilon$. We assume the wavefunctions of the two
states to be equal, that leads to an error proportional 
to $(\as/m_b^2)$~\cite{JA}.
Taking the ratio of the eqs.~(\ref{eq:widsc1l}) and~(\ref{eq:widvec1l}) 
and expanding to first order in $\as$, we obtain:
\beqy
\frac{\Gamma (\eta_b\rightarrow \gamma\gamma)}{\Gamma(\Upsilon
\rightarrow e^+e^-)}\approx
\frac{1}{3} \frac{(1-3.38\as/\pi)}{(1-5.34\as/\pi)} = \nonumber \\
= \frac{1}{3}  \left [ 1+  1.96 \frac{\alpha_s}{\pi} 
 + \mathcal{O}(\alpha_s^2) \right ] .
\label{eq:ratio1loop}
\eeqy
In order to compute the correction we start from the two-loop expression 
for $\as$ and the value at the $Z$ 
mass~\cite{PDG} $\alpha_s(M_Z) = 0.118\pm 0.003 $. 
Using the  renormalization group equation to evaluate 
$\alpha_s(Q=2m_b=10.0\mbox{ GeV})=0.178 \pm 0.007$, and the  latest 
measurement of the $\Upsilon$
\be
 \Gamma_{exp}(\Upsilon \to e^+e^-) = 1.32 \pm 0.05 \textrm{ keV }
\label{eq:upsexpwid}
\ee
for the $\etab$ one obtains 
\be
\Gamma(\etab\rightarrow \gamma\gamma) \pm
\Delta\Gamma(\etab\rightarrow \gamma\gamma) =489 \pm 19 \pm 2 \textrm{ eV ,}
\label{eq:singnaive}
\ee
where the first error comes from the uncertainty on the $\Upsilon$ experimental
width, the second error from $\alpha_s$~.
We have assumed the $\as$ scale to be $Q=2m_b=10.0\textrm{ GeV}$. This
choice is by no way unique as shown for the $\etac$ decay~\cite{NOI2},
 and in fig.~(\ref{fig:oneloop}) we show the 
dependence of the $\etab$ photonic  width, evaluated  from 
eq.~(\ref{eq:upsexpwid}), upon  different values of the scale 
chosen for $\as$~.

Unlike  the $\etac$ case there are no experimental measurements
of this decay; we shall assume therefore that it is not 
possible to determine a scale choice of $\as$ for the $\etab$ decay.
We will have to include this fluctuation in the indetermination due to 
radiative corrections.

\section{Potential Model predictions}
We present now results for $\etab \to \gamma \gamma$ from the potential models.
This method allows us to obtain the absolute width, through   
the wave function at the origin computed from a Schr\"odinger equation.
For the calculation of the wavefunction~\cite{SCHROEDINGER}
we have used four different 
potential models, starting from the one of Rosner et al.~\cite{ROSNER}
$ V(r) = \lambda ((r/r_0)^{\alpha} - 1 )/\alpha +C \mbox{ .}$
with $r_0 = 1 \textrm{ GeV}^{-1} , \;  \alpha=-0.14 \mbox{ , } 
\lambda=0.808\textrm{ GeV, } C=-1.305\textrm{ GeV}$, 
and  the QCD inspired potential $V_{J}$  of Igi-Ono \cite{IGI,TYE}
\beqy
 V_{J}(r) = V_{AR}(r) + d r e^{-gr} + ar , \nonumber \\
 V _{AR}(r) = -\frac{4}{3} \frac{\alpha_{s}^{(2)}(r)}{r} \label{eq:igipot}
\eeqy 
with two different  parameter sets, corresponding to 
 $\Lambda_{\overline{MS}}=0.5 \  GeV$
and $\Lambda_{\overline{MS}}=0.2 \  GeV$ respectively \cite{IGI}.
We show also the results from a Coulombic type potential with the QCD 
coupling $\alpha_s$ frozen
to a value of $r$ which corresponds to the Bohr radius of the
quarkonium system, $r_B=3/(2m_b\as(r_B))$ (see for instance \cite{NOI1}).
The latter has the advantage of providing analytical solutions for
the wavefunction and the energy levels:
$ E_n = -  4m\as(r_B)^2/(9n^2)$ and
$ |\psi(0)|^2 =   ( 2m\as(r_B)/3  )^3/\pi $

\begin{figure}[h]
\begin{center}
\includegraphics*[angle=0,width=0.4\textwidth]{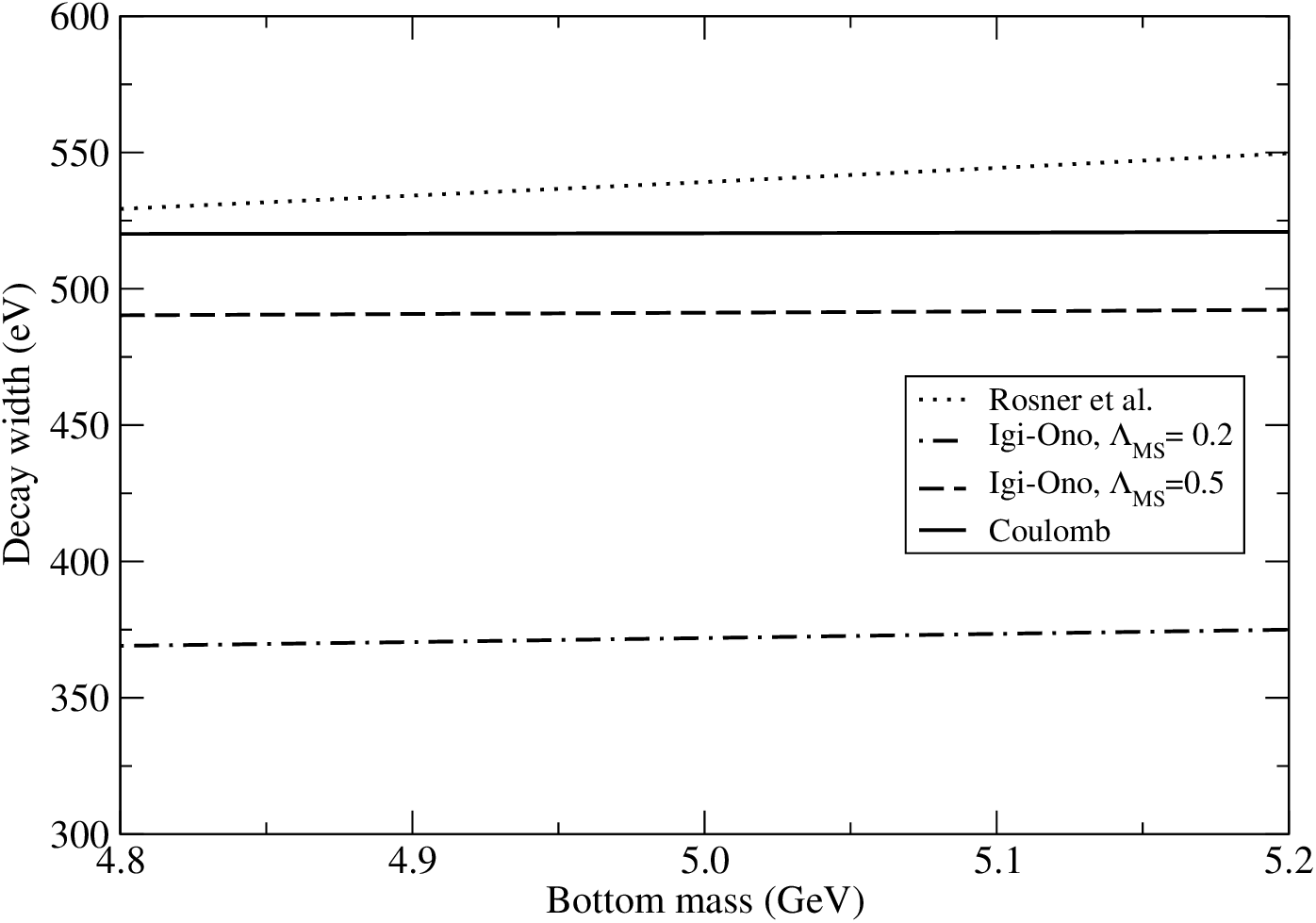}  
\caption{The dependence of $\etab$ decay width to $\gamma\gamma$ in 
\emph{eV} for  different potential models is shown as a 
function of $m_b$. The perturbative correction scale is here
set to be $Q=2m_b$.\label{fig:potentials}}
\end{center}
\end{figure}

We have to stress the fact that the scale of $\as$ occurring in the radiative 
correction and the one of Coulombic potential are different.
We show in fig.~(\ref{fig:potentials}) the predictions for the decay width 
from these  potential models with the correction from 
eq.~(\ref{eq:widsc1l}) at an $\as$ scale $Q=2m_b$~.

For any given model, sources of error in this calculation arise from the
choice of scale in the radiative correction factor discussed before and 
the choice of the parameters. The difference of the prediction between
the various models introduce another indetermination for the decay value.
We can estimate a range of values for the  potential model predictions for 
the radiative decay width  $\Gamma(\etab \to \gamma \gamma)$,
namely:

\be
\Gamma(\etab \to \gamma \gamma) =466 \pm 101 \textrm{ eV \mbox{       },} 
\label{eq:potmodpred}
\ee
that is an indetermination of around 22\%.

\section{Octet component procedure}
We will  present now another approach which admits other components to the 
meson decay beyond the one from the colour singlet picture (Bodwin, 
Braaten and Lepage)~\cite{BBL}. 
NRQCD has been used to separate the short distance scale of 
annihilation from the nonperturbative contributions of long distance scale.
This model has been successfully used to explain the larger than expected 
$\jpsi$ production at the Tevatron and LEP.
The decay width expression for a pseudoscalar $P$ or a vector state $V$  is 
given by means of NRQCD from the expansion:
\be
\Gamma_{V,P} = \sum_{n} \frac{2\Im f_n(\as)}{m_q^{d{_n}-4}} 
\langle \mathcal{O}_n \rangle_{V,P}
\ee
that is a sum of terms in inverse powers of $m_q$, each of which factorises 
into a perturbative coefficient 
$\Im f_n(\as)$ and  nonperturbative matrix element 
$\langle \mathcal{O}_n\rangle_{V,P}$. 

According to BBL, in the octet model for quarkonium, the electromagnetic 
and light hadrons ($LH$) decay widths of bottomonium states are given by:

$$\Gamma(\Upsilon \to LH) = 
\frac{2\langle \Upsilon | O_1(^3S_1) | \Upsilon \rangle}{m_b^2}  
\left (\frac{10}{243}\pi^2- \frac{10}{27}\right) \as^3 $$

$$\times  \left [ 1+\left (-9.46 \times  \frac{4}{3}+ 12.39-1.161n_f 
\right ) \frac{\as}{\pi} \right ] + $$

\be
\frac{2\langle \Upsilon | P_1(^3S_1) | \Upsilon \rangle}{m_b^4}
 \frac{17.32\times \left [20 (\pi^2-9 )\right ]}{486} \as^3 
\label{eq:bblups2lh} 
\ee

$$\Gamma(\Upsilon \to e^+e^-) = 
\frac{2\langle \Upsilon | O_1(^3S_1) | \Upsilon \rangle}{m_b^2}  
\left [ \frac{\pi}{3}Q^2\alpha^2 \left (1- \right . \right .$$
\be
\left . \left . \frac{13}{3}\frac{\as}{\pi}\right)\right ] -
\frac{2\langle \Upsilon | P_1(^3S_1) | \Upsilon \rangle}{m_b^4}\frac{4}{9}\pi
 Q^2\alpha^2
\label{eq:bblups2ee}
\ee

$$
\Gamma(\etab \to LH) = \frac{2\langle \etab 
| O_1(^1S_0) | \etab \rangle}{m_b^2} 
\frac{2}{9} \pi \as^2 \left [1+\left (\frac{53}{2} \right . \right .$$
\be
\left . \left .-\frac{31}{24}
\pi^2-\frac{8}{9} n_f \right ) \right ] 
-\frac{2\langle \etab | P_1(^1S_0) | \etab \rangle}{m_b^4}\frac{8}{27}\pi
 \as^2
 \label{eq:bbleta2lh}
\ee

$$
\Gamma(\etab \to \gamma \gamma) = \frac{2\langle \etab 
| O_1(^1S_0) | \etab \rangle}{m_b^2} \pi Q^4 \aqed^2 
\left [1+ \right . $$

\be
\left . \left (\frac{\pi^2-20}{3}\right )\frac{\as}{\pi} \right ]-
 \frac{2\langle \etab | P_1(^1S_0) | \etab \rangle}{m_b^4} 
\frac{4}{3} \pi Q^4\aqed^2
\label{eq:bbleta2gamgam}
\ee

There are four unknown long distance coefficients, which can be reduced to
two by means of the vacuum saturation approximation:
\be
G_1 \equiv \langle \Upsilon | O_1(^3S_1) | \Upsilon \rangle = \langle \etab 
| O_1(^1S_0) | \etab \rangle
\ee
\be
F_1 \equiv \langle \Upsilon | P_1(^3S_1) | \Upsilon \rangle = \langle \etab 
| P_1(^1S_0) | \etab \rangle
\ee
which is correct up to $\mathcal{O}(v^2)$, where ${\vec{v}}$ is  the quark
velocity inside the meson.
In the potential model language this translates to consider the two
wavefunctions of the pseudoscalar and the vector state to be equal.
With this position we obtain a system of equations for the nonperturbative
coefficients $G_1$ and $F_1$ of
$\Upsilon$ electromagnetic decay  and light hadrons widths which in turn
allows us to compute the $\etab$ decay widths.

The BBL approach gives the following decay widths of the $\etab$ meson:

\be
\Gamma(\etab\rightarrow \gamma\gamma)  =364 \pm 8 \pm 13 \textrm{ eV}
\label{eq:bblres2gamma}
\ee
and 
\be
\Gamma(\etab\rightarrow LH)  =57.9 \pm 4.6 \pm 2.8 \textrm{ keV ,}
\ee
where the first error comes from the uncertainty on the $\Upsilon$ 
experimental width, the second error from $\alpha_s$~.
The improvement of the error on eq.~(\ref{eq:bblres2gamma})
with respect to the previous analogous
determination on the $\eta_c$ decay~\cite{NOI2} is due to  better error on the
experimental measures of the $\Upsilon$ decay widths compared to the one 
of the $J/\psi$, and the smaller indetermination on the $\as$ value due to
the higher energy scale involved in the decay. These reasons, together with 
the fact that the potential models used are fitted for the $c\overline{c}$  
system, justifies the improvement of accuracy given in 
eq.~(\ref{eq:bblres2gamma}) compared to the one of 
eq.~(\ref{eq:potmodpred}).

\section{Comparison between models}

\begin{figure}[h]
\begin{center}
\includegraphics*[angle=0,width=0.4\textwidth]{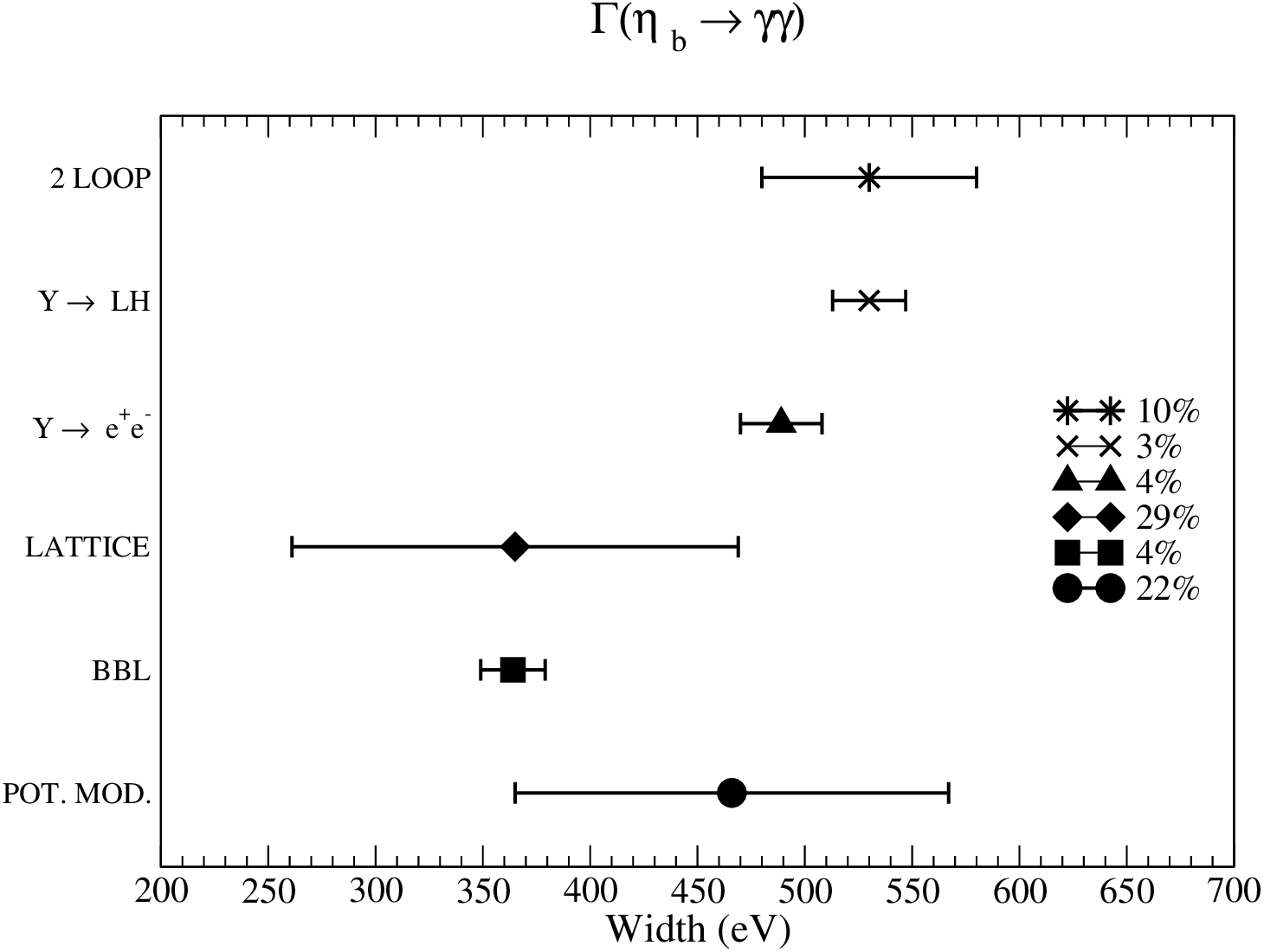}
\caption{ The $\etab$ two photon width as calculated in this paper 
using (starting from below) Potential Models results, BBL procedure with 
input from $J/\psi$ decay data, Lattice evaluation of
$G_1$ and $F_1$ factors, Singlet picture with $G_1$ obtained from 
$\Upsilon \to e^+e^-$ and $\Upsilon \to LH$ processes 
respectively, and the two--loop enhanced procedure.
The legend shows the percentage error on each value.
\label{fig:summary}}
\end{center}
\end{figure}

For comparison we present in fig.~(\ref{fig:summary})
a set of predictions coming from different methods. 

Starting with  potential models, we see that the results are in good
agreement with each other. The advantage of this method is that we are 
giving a prediction from first principles,  without using any 
experimental input. Since there are currently no experimental measures for the
$\etab \to \gamma \gamma$ decay, we shall use this prediction as a 
reference point, as it has proven to be reliable in the case of 
charmonium decay~\cite{NOI2}.
The second evaluation, given by BBL 
using the experimental values of the $\Upsilon$ decay, is on the left limit 
of the potential models value. This is true also for the determination of 
the BBL procedure  with nonperturbative long distance terms taken from
from the lattice calculation~\cite{LATTICE}, affected from a large error
of around 30\%. 
The advantage of the latter is that its prediction, like the one from 
potential models, does not make use of any experimental value. 
Next is the point given by the singlet picture from the electromagnetic 
decay of the $\Upsilon$,  aligned with the aforementioned results of the BBL
procedure. 
The point above is obtained also from  the singlet picture with the $\Upsilon$ 
decay into light hadrons, in agreement with the results given from the potential
models. 
We notice that in analogy to the charmonium case (see~\cite{NOI2} and references
therein) the singlet results obtained from the $\Upsilon$ decay are in 
disagreement with each other, in this case by only $1 \sigma$.
The last point from a two--loop enhanced calculation 
given by~\cite{CZARNECKI,ALEPH} is in agreement with the potential model 
result and the singlet decay from the $\Upsilon \to LH$ process.

\section{Conclusions}
The $\Gamma(\etab \to \gamma \gamma)$ decay width prediction of the 
potential models considered gives the value $ 466 \pm 101  \textrm{ eV}$, in
agreement with the naive estimate from the $\Upsilon$ decay given 
by~(\ref{eq:singnaive}). Predictions of the BBL procedure are consistent with
the potential model results, for both the long distance terms $G_1$ and $F_1$ 
extracted from the $\Upsilon$ experimental decay widths and the one evaluated
from lattice calculations. The results from the singlet picture are also
consistent with the potential model results. Finally the two--loop enhanced 
prediction is in good agreement with the potential model results.


\begin{thebibliography}{9}

\bibitem{ALEPH}

A. B\"ohrer, hep-ph/0110030, to appear in proceedings of
``PHOTON 2001'', Ascona, Switzerland (2001), ed. by M. Kienzle, World Sci.,
Singapore, 2001, and private communication;
The ALEPH Collaboration, A. Heister et al., Phys. Lett.B 530 (2002) 56.
\bibitem{JA} Nicola Fabiano
Eur. Phys. J. C26 (2003) 441-444;
N.~Fabiano , G.~Pancheri,
To appear in the proceedings of ``1st International Workshop on 
Frontier Science: Charm, Beauty, and CP'', Frascati, Rome, Italy, 6-11 Oct 2002.
e-Print Archive: hep-ph/0210279.
\bibitem{CDF2K2}
CDF Collaboration (J.~Tseng for the collaboration), FERMILAB-Conf-02/348-E.
\bibitem{NOI2}
N. Fabiano, G. Pancheri, hep-ph/0110211,
 to appear in proceedings of ``PHOTON 2001'', Ascona, Switzerland 
(2001), ed. by M. Kienzle, World Sci., Singapore, 2001; 
N. Fabiano, G. Pancheri, Eur. Phys. J. C25 (2002) 421. 

\bibitem{ROSNER}
A.K. Grant, J.L. Rosner and E. Rynes, Phys. Rev. D47  (1993) 1981.

\bibitem{IGI}
J. H. K\"uhn and S. Ono, Zeit. Phys.C21
(1984) 385;
K. Igi and S. Ono, Phys. Rev. D33 (1986) 3349.

\bibitem{NOI1}
N. Fabiano, A. Grau and G. Pancheri, Phys. Rev. D50 (1994) 
3173;  Nuovo Cimento A, Vol 107  (1994) 2789.

\bibitem{BBL}
G.T. Bodwin, E. Braaten and G.P. Lepage, Phys. Rev.~D51 (1995) 1125.


\bibitem{CZARNECKI}
A. Czarnecki and K. Melnikov, Phys. Lett.B 519 (2001) 212.

\bibitem{VANROYEN}
R.Van Royen and V.Weisskopf, Nuovo Cimento 50A (1967) 617.

\bibitem{BARBIERI}
R. Barbieri, G. Curci, E. d'Emilio and R. Remiddi
Nucl. Phys. B154 (1979) 535.

\bibitem{MACKENZIE}
P. Mackenzie and G. Lepage, Phys. Rev. Lett.47 (1981) 1244.

\bibitem{CORNELL}
E. Eichten, K. Gottfried, T. Kinoshita, K. D. Lane and T. M. Yan,
Phys. Rev. 21D (1980) 203.

\bibitem{EICHTEN}
E. Eichten and F. Feinberg, Phys. Rev. D23, (1981) 2724.


\bibitem{PDG}
Review of Particle Properties, K. Hagiwara et al., Phys. Rev. D66, (2002)
010001; http://pdg.lbl.gov/.



\bibitem{SCHROEDINGER}
Thanks to F.F.~Sch\"obrl for providing the program; 
W.~Lucha, F.F.~Sch\"obrl, Int. J. Mod. Phys. C 10 (1999) 607.

\bibitem{TYE}
W. Buchmuller and S. H. H. Tye, Phys. Rev. D24 (1981) 132.

\bibitem{LATTICE}
G.T. Bodwin, D.K. Sinclair and S. Kim, 
Int. J. Mod. Phys. A 999 (2001) 123.



\end{thebibliography}
\end{document}